\documentclass[12pt, a4paper, oneside, openany]{article}

\usepackage[a4paper, lmargin=0.1\paperwidth, rmargin=0.17\paperwidth, tmargin=0.1111\paperheight, bmargin=0.1111\paperheight]{geometry} 

\usepackage[english]{babel}
\usepackage{parskip}
\usepackage{rotating}
\usepackage{amssymb,mathtools}
\usepackage{helvet} 
\usepackage{graphicx}
\usepackage{tabularx, booktabs, multirow}
\usepackage{ragged2e}
\newcolumntype{R}{>{\RaggedRight\arraybackslash}X}
\usepackage[justification=raggedright]{caption}
\usepackage[stable,hang,flushmargin]{footmisc} 
\usepackage{xcolor}
\usepackage{csquotes} 
\usepackage[pages=some]{background}
\usepackage[sectionbib]{natbib}
\usepackage{chapterbib} 

\usepackage{longtable}
\usepackage{graphicx,dcolumn}
\usepackage{algorithm}
\usepackage{algorithmic}
\usepackage[inline]{enumitem}
\usepackage{svg}
\usepackage{float}
\usepackage{array}
\usepackage{xurl}
\usepackage{mathptmx}%
\usepackage{ltablex}
\usepackage{makecell}
\usepackage[caption=false]{subfig}
\usepackage{tablefootnote}
\newcommand{\ra}[1]{\renewcommand{\arraystretch}{#1}}
\usepackage{acro}
\DeclareAcronym{DEFI}{
  short = DeFi,
  long  = decentralised finance,
  long-plural-form = Decentralised Finances
}
\DeclareAcronym{VC}{
  short = VC,
  long  = venture capital
}
\DeclareAcronym{DAO}{
  short = DAO,
  long  = decentralised autonomous organisation
}
\DeclareAcronym{CME}{
  short = CME,
  long = Chicago Mercantile Exchange
}
\DeclareAcronym{CBOE}{
  short = CBOE,
  long = Chicago Board Options Exchange
}
\DeclareAcronym{LOIH}{
  short = LOIH,
  long = Large Open Interest Holders
}
\DeclareAcronym{FCM}{
  short = FCM,
  long = Futures Commission Merchant
}
\DeclareAcronym{GBTC}{
  short = GBTC,
  long = Grayscale Bitcoin Trust
}
\DeclareAcronym{ETF}{
  short = ETF,
  long = exchange traded fund
}
\DeclareAcronym{SEC}{
  short = SEC,
  long = Securities and Exchange Commission
}
\DeclareAcronym{EIOPA}{
  short = EIOPA,
  long = European Insurance and Occupational Pensions Authority
}
\DeclareAcronym{MiCA}{
  short = MiCA,
  long = markets in crypto assets
}
\DeclareAcronym{SMSF}{
  short = SMSF,
  long = self-managed superannuation fund
}

\DeclareAcronym{ICO}{
  short = ICO,
  long = Initial Coin Offerings
}
\DeclareAcronym{CFTC}{
  short = CFTC,
  long = Commodity Futures Trading Commission
}
\DeclareAcronym{IRA}{
  short = IRA,
  long = Individual Retirement Account
}
\DeclareAcronym{TIP}{
  short = TIP,
  long = Teachers' Innovation Platform
}
\DeclareAcronym{IRS}{
  short = IRS,
  long = Internal Revenue Service
}
\DeclareAcronym{TVL}{
  short = TVL,
  long = Total Value Locked
}
\DeclareAcronym{FSB}{
  short = FSB,
  long = Financial Stability Board
}
\DeclareAcronym{NSA}{
  short = NSA,
  long = National Security Agency
}
\DeclareAcronym{LLC}{
  short = LLC,
  long = Limited Liability Company
}
\DeclareAcronym{SPoF}{
  short = SPoF,
  long = Single Point of Failure
}
\DeclareAcronym{FATF}{
  short = FATF,
  long = Financial Action Task Force
}
\DeclareAcronym{FinCen}{
  short = FinCen,
  long = Financial Crimes Enforcement Network
}
\DeclareAcronym{SA}{
  short = SA,
  long = Securities Act
}
\DeclareAcronym{SEA}{
  short = SEA,
  long = Securities Exchange Act
}
\DeclareAcronym{MICA}{
  short = MICA,
  long = Markets in Crypto-Assets
}
\DeclareAcronym{CEA}{
  short = CEA,
  long = Commodity Exchange Act
}
\DeclareAcronym{DLT}{
  short = DLT,
  long = distributed ledger technology,
  long-plural-form = distributed ledger technologies
}
\DeclareAcronym{EBA}{
  short = EBA,
  long = European Banking Authority
}
\DeclareAcronym{FCA}{
  short = FCA,
  long = Financial Conduct Authority
}
\DeclareAcronym{CBDC}{
  short = CBDC,
  long = Central Bank Digital Currency
}
\DeclareAcronym{CPI}{
  short = CPI,
  long = Consumer price indices
}
\DeclareAcronym{AML}{
  short = AML,
  long = Anti-money laundering
}
\DeclareAcronym{KYC}{
  short = KYC,
  long = Know Your Customer
} 
\DeclareAcronym{MLR}{
  short = MLR,
  long = {Money Laundering, Terrorist Financing and Transfer of Funds (Information on the Payer) Regulation}
} 
\DeclareAcronym{TradFi}{
    short = TradFi,
    long = Traditional Finance
}
\DeclareAcronym{CDD}{
  short = CDD,
  long = customer due diligence 
} 
\DeclareAcronym{EDD}{
  short = EDD,
  long = enhanced customer due diligence 
}
\DeclareAcronym{POCA}{
  short = POCA,
  long = Proceeds of Crime Act 
} 
\DeclareAcronym{DPA}{
  short = DPA,
  long = Data Protection Act
} 
\DeclareAcronym{DApp}{
  short = DApp,
  long = Decentralised Application
} 

\DeclareAcronym{TACT}{
  short = TACT,
  long = Terrorism Act
} 
\DeclareAcronym{SAR}{
  short = SAR,
  long = suspicious activity report
} 
\DeclareAcronym{JMLSG}{
  short = JMLSG,
  long = Joint Money Laundering Steering Group
} 
\DeclareAcronym{VA}{
  short = VA,
  long = virtual asset
}
\DeclareAcronym{VASP}{
  short = VASP,
  long = virtual asset service provider
}
\DeclareAcronym{ATH}{
    short = ATH,
    long = all time high
}
\DeclareAcronym{5MLD}{
  short = 5MLD,
  long = Fifth Money Laundering Directive
} 
\DeclareAcronym{ATM}{
  short = ATM,
  long = automated teller machines
} 
\DeclareAcronym{FIU}{
  short = FIU,
  long = financial intelligence unit
}
\DeclareAcronym{NCA}{
  short = NCA,
  long = National Crime Agency
}
\DeclareAcronym{FinCEN}{
  short = FinCEN,
  long = Financial Crimes Enforcement Network
}
\DeclareAcronym{RBA}{
  short = RBA,
  long = risk-based approach
}

\DeclareAcronym{gsc}{
  short = GSC,
  long = global stable coin
}

\DeclareAcronym{cpmi}{
  short = CPMI,
  long = the Committee on Payments and Market Infrastructures
}

\DeclareAcronym{ecb}{
  short = ECB,
  long  = The European Central Bank
}
\DeclareAcronym{boe}{
  short = BoE,
  long  = The Bank of England
}

\DeclareAcronym{ccp}{
  short = CCP,
  long = central counterparty
}

\DeclareAcronym{iosco}{
  short =  IOSCO,
  long = International Organization of Securities Commissions
}

\DeclareAcronym{OOS}{
  short =  OOS,
  long = out-of-sample
}

\DeclareAcronym{REIT}{
  short =  REIT,
  long = public real estate
}

\DeclareAcronym{DMSPE}{
  short =  DMSPE,
  long = the discounted mean squared prediction error
}

\DeclareAcronym{FRED}{
  short =  FRED,
  long = Federal Reserve Economic Data
}

\DeclareAcronym{SR}{
  short =  SR,
  long = Sharpe ratio
}

\DeclareAcronym{CeFi}{
  short =  CeFi,
  long = Centralised Finance
}

\DeclareAcronym{CSA}{
  short =  CSA,
  long =  credit support annex
}

\DeclareAcronym{PoW}{
  short =  PoW,
  long =  Proof-of-Work
}

\DeclareAcronym{PoS}{
  short =  PoS,
  long =  Proof-of-Stake
}

\DeclareAcronym{DEX}{
  short =  DEX,
  long =  decentralised exchange
}

\DeclareAcronym{LP}{
  short =  LP,
  long =  liquidity provider
}
\DeclareAcronym{PLF}{
    short = PLF,
    long = protocol for loanable funds,
    long-plural-form = protocols for loanable funds
}

\DeclareAcronym{LT}{
  short =  LT,
  long =  liquidity taker
}

\DeclareAcronym{AMM}{
  short =  AMM,
  long =  Automated Market Maker
}

\DeclareAcronym{HyFi}{
  short =  HyFi,
  long =  Hybrid Finance
}
\DeclareAcronym{CPMM}{
  short =  CPMM,
  long =  Constant Product Market Maker
}

\DeclareAcronym{MEV}{
  short =  MEV,
  long =  miner extractable value
}

\DeclareAcronym{PDR}{
  short =  PDR,
  long =  Protocol Defined Return
}
  
\DeclareAcronym{TEA}{
  short = TEA,
  long =  triple-entry accounting
}  

\DeclareAcronym{TEB}{
  short = TEB,
  long =  triple-entry book-keeping
} 

\DeclareAcronym{STRs}{
  short = STRs,
  long =  Shared Transaction Repositories 
}  
 
\DeclareAcronym{DBT}{
  short = DBT,
  long =  Distributed Book Technology
}  
 
\DeclareAcronym{DJT}{
  short = DJT,
  long =  Distributed Journal Technology
} 
  
\DeclareAcronym{P2P}{
  short = P2P,
  long =  peer-to-peer
}
\DeclareAcronym{GLT}{
  short = GLT,
  long =  General Ledger for Transactions
}

\DeclareAcronym{GLR}{
  short = GLR,
  long =  General Ledger for Reporting
}

\DeclareAcronym{IDEA}{
  short = IDEA,
  long =  Immutable Double-entry Accounting
}

\DeclareAcronym{PTA}{
  short = PTA,
  long =  Plain Text Accounting
}

\DeclareAcronym{NSL}{
  short = NSL,
  long =  National security letter
}

\DeclareAcronym{RFPA}{
  short = RFPA,
  long =  Right to Financial Privacy Act
}

\DeclareAcronym{ECPA}{
  short = ECPA,
  long =  Electronic Communications Privacy Act
}

\DeclareAcronym{SACs}{
  short = SACs,
  long =  Special Agents in Charge 
}

\DeclareAcronym{AECs}{
  short = AECs,
  long =  Anonymity Enhanced Cryptocurrencies
}

\DeclareAcronym{MSBs}{
  short = MSBs,
  long =  money service businesses
}

\DeclareAcronym{BIS}{
  short = BIS,
  long =  International Settlements
}
\DeclareAcronym{IMF}{
  short = IMF,
  long =  International Monetary Fund
}

\DeclareAcronym{DCF}{
    short = DCF,
    long = discounted cash flow
}

\DeclareAcronym{EMH}{
    short = EMH,
    long = Efficient Market Hypothesis
}

\DeclareAcronym{GGM}{
    short = GGM,
    long = Gordon Growth Model
}

\DeclareAcronym{UNI}{
    short = UNI,
    long = Uniswap
}

\DeclareAcronym{CRV}{
    short = CRV,
    long = Curve
}

\DeclareAcronym{ICE}{
    short = ICE,
    long = Intercontinental Exchange
}

\DeclareAcronym{NDAQ}{
    short = NDAQ,
    long = Nasdaq
}

\DeclareAcronym{CRSP}{
    short = CRSP,
    long = Center for Research in Security Prices
}

\DeclareAcronym{WACC}{
    short = WACC,
    long = Weighted Average Cost of Capital
}

\DeclareAcronym{GDP}{
    short = GDP,
    long = Gross Domestic Product
}

\DeclareAcronym{CQGR}{
    short = CQGR,
    long = Compound Quarter Growth Rate
}

\DeclareAcronym{AAVE}{
    short = AAVE,
    long = Aave
}

\DeclareAcronym{YFI}{
    short = YFI,
    long = Yearn Finance
}

\DeclareAcronym{IDLE}{
    short = IDLE,
    long = Idle Finance
}

\DeclareAcronym{BRK.B}{
    short = BRK.B,
    long = Berkshire Hathaway
}

\DeclareAcronym{COMP}{
    short = COMP,
    long = Compound
}

\DeclareAcronym{C}{
    short = C,
    long = Citigroup
}

\DeclareAcronym{BAC}{
    short = BAC,
    long = Bank of America
}

\DeclareAcronym{WFC}{
    short = WFC,
    long = Wells Fargo \& Co.
}

\DeclareAcronym{MS}{
    short = MS,
    long = Morgan Stanley
}

\DeclareAcronym{BLK}{
    short = BLK,
    long = BlackRock
}

\DeclareAcronym{CEX}{
    short = CEX,
    long = Centralised Exchange
}

\DeclareAcronym{NPV}{
    short = NPV,
    long = Net Present Value
}

\PassOptionsToPackage{hyphens}{url}
\usepackage{hyperref} 
\hypersetup{colorlinks=true,
	linkcolor=blue,
	citecolor=blue,
	urlcolor=blue}

\addto\extrasenglish{%
}
\addto\extrasenglish{%
}	

\addto\extrasenglish{%
}

\usepackage[clearempty]{titlesec}

\newcommand{\chapterbib}{%
\bibliographystyle{apalike}
\bibliography{references}%
}

\begin{document}
\title{DeFi versus TradFi: Valuation Using Multiples and Discounted Cash Flows}
\date{}
\maketitle
\vspace{-1cm}

\begin{center}
\author{%
  Teng Andrea Xu\footnote{AQR Capital Management}%
  \and\hspace{0.25cm}Jiahua Xu\footnote{Centre for Blockchain Technologies, University College of London and Exponential Science}%
  \and\hspace{0.25cm}Kristof Lommers\footnote{University of Oxford}%
  }
\end{center} 

\vspace{1.5
cm}

\begin{abstract}
As of August 2022, blockchain-based assets boast a combined market capitalisation exceeding one trillion USD, among which the most prominent are the \acf{DAO}\index{DAO} tokens associated with \ac{DEFI}\index{DeFi} protocols. In this work, we seek to value \ac{DEFI}\index{DeFi} tokens using the canonical multiples and \ac{DCF} approaches. 
We examine a subset of \ac{DEFI}\index{DeFi} services including \acp{DEX}\index{DEX}, \acp{PLF}, and yield aggregators\index{aggregator}. 
We apply the same analysis to some publicly traded firms and compare them with \ac{DEFI}\index{DeFi} tokens of the analogous category. 
Interestingly, despite the crypto bear market lasting for more than one year as of August 2022, both approaches evidence overvaluation in \ac{DEFI}\index{DeFi}.
\end{abstract}

\section{Introduction}

\begin{figure}[H]
    \centering
    \subfloat[ \label{fig:dex} \acp{DEX}.]{
        \includegraphics[width=0.3\linewidth]{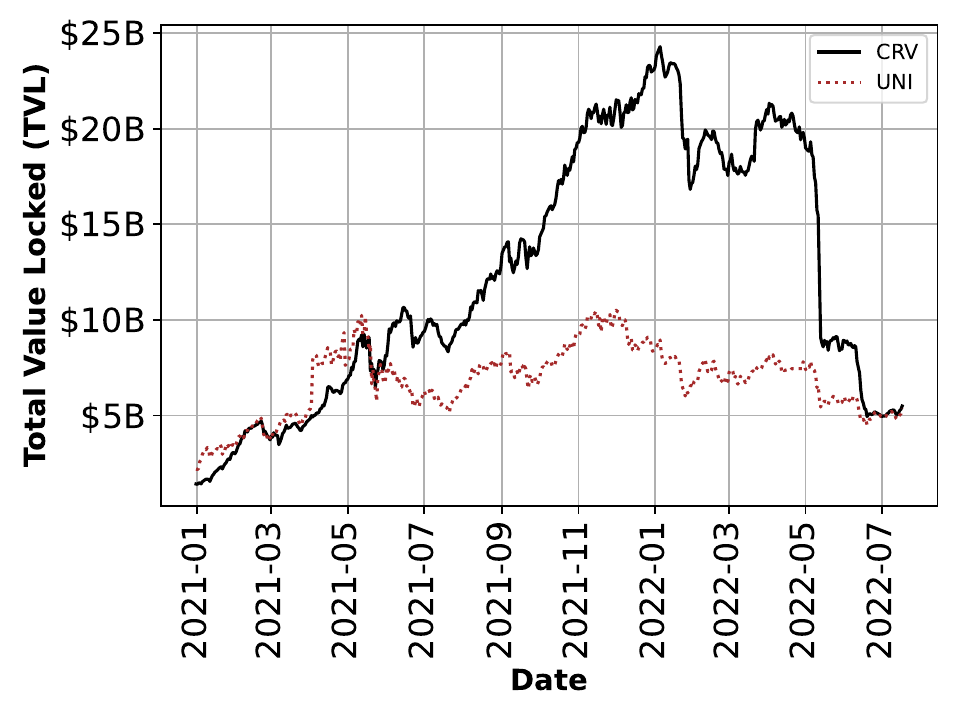}
    }
    \subfloat[ \label{fig:plf} \acp{PLF}.]{
        \includegraphics[width=0.3\linewidth]{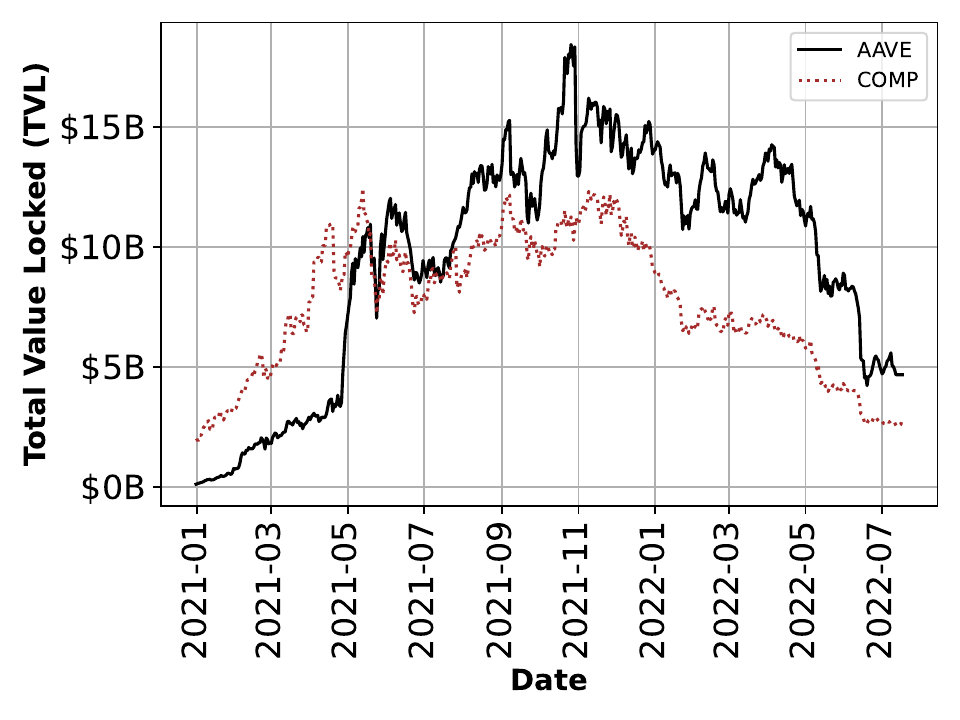}
    }
    \subfloat[\label{fig:yield_agg} Yield aggregators.]{
        \includegraphics[width=0.3\linewidth]{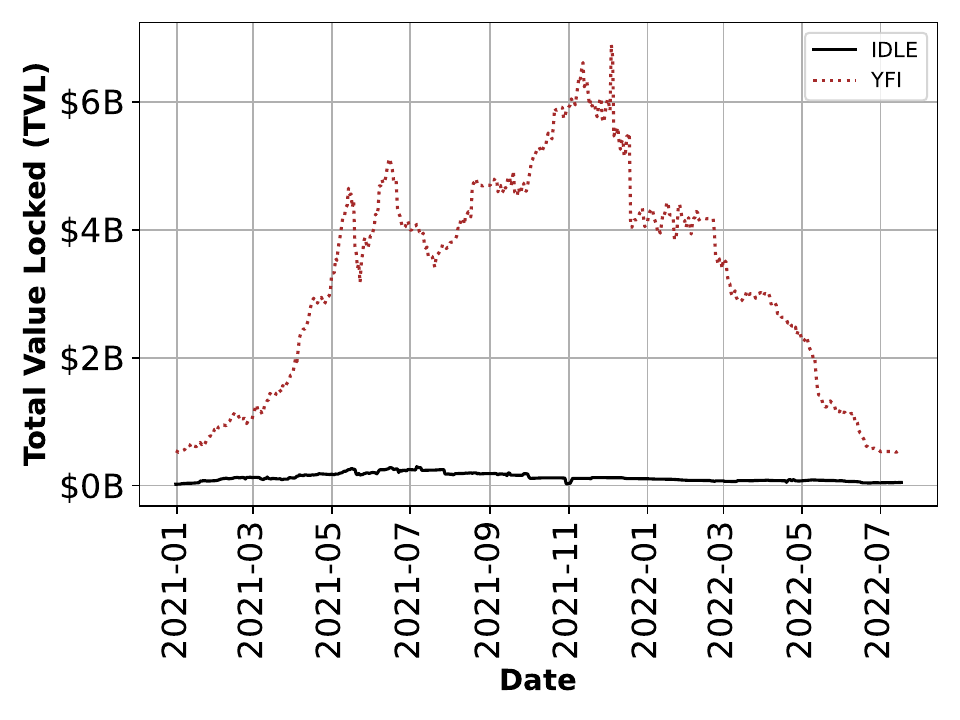}
    }
    \caption{The figure above shows the historical \ac{TVL} for different \ac{DEFI}\index{DeFi}.} 
    \label{fig:total_value_locked}
\end{figure}

\ac{DEFI}\index{DeFi} provides financial services such as loans or asset exchange\index{exchange} leveraging \acp{DLT}. 
Towards the end of summer 2020, the total market cap of \ac{DAO}\index{DAO} tokens in \ac{DEFI}\index{DeFi} grew to 154 billion USD from 1.8 billion USD as of March 2020 \citep{Xu2023AProtocols}. 
The \ac{TVL}, that is, the net worth of all assets locked in smart contracts, reached its peak for many \ac{DEFI}\index{DeFi} protocols in that period (\autoref{fig:total_value_locked}).

Since that \enquote{\ac{DEFI}\index{DeFi} Summer}, \ac{DAO}\index{DAO} tokens connected to decentralised financial services have enjoyed a large growth of interest both within the crypto community and beyond, despite the continuous debate surrounding whether \ac{DEFI}\index{DeFi} is just a fad, or is to stay and co-exist with \ac{TradFi}\index{TradFi} \citep{qin2021cefi}.
This interest was mainly driven from \ac{DEFI}\index{DeFi}'s \textit{permissionless} \citep{Werner2022SoK:DeFi} and \textit{composable} nature \citep{qin2021cefi}. 

The growth of interest has brought more users in the protocols with
\begin{enumerate*}[label={(\roman*)}]
\item higher on-chain transaction volume\index{volume} and
\item higher off-chain trading volume\index{volume}.
\end{enumerate*}
Similarly, entrepreneurs and hedge funds have decided to invest into digital assets, such as Grayscale and BlackRock, to cite a few.
However, as a myriad of \ac{DEFI}\index{DeFi} projects provide similar financial services, it is hard to reason why an investor should prefer one over the other. 
In classical finance, investors use valuation theory \citep{damodaran2007valuation} for their decision-making process in asset selection. 

On the one hand, the necessity for fundamental analysis is the fact that market prices often do not reflect the true underlying value of an asset.
Fundamental analyses look into firms' financial statements and estimate a {\em fair} share price based on the firm's performance. 
On the other hand, comparable analysis assumes that, on average, the market prices reflect investors' beliefs \citep{damodaran2007valuation}. 
Comparable analysis will still look at financial fundamentals, but instead of estimating a theoretical value, it will use firms' fundamental value ratios, that is, \enquote{multiples} to compare them and thus spot comparatively undervalued/overvalued assets. Naturally, comparable analysis is only applicable to firms within a similar segment.

In this study, we apply both
intrinsic and
relative valuation methods to selected \ac{DAO}\index{DAO} tokens in \ac{DEFI}\index{DeFi} as well as stocks in \ac{TradFi}\index{TradFi}.
Our contributions can be summarised as follows:
\begin{enumerate}
    \item First, we apply empirically fundamental and comparable analyses to \ac{DEFI}\index{DeFi} \ac{DAO}\index{DAO} tokens. Surprisingly, although \ac{DEFI}\index{DeFi} tokens are in a bear market for more than 1 year thus far, they remain overvalued compared to their theoretical fair value.
    \item Second, we compare multiples between \ac{DEFI}\index{DeFi} and \ac{TradFi}\index{TradFi}. This comparison shows that \ac{DEFI}\index{DeFi} tokens multiples exceed their \ac{TradFi}\index{TradFi} counterparties by several times, demonstrating the general overvaluation of the former.
    \item Finally, all analyses use publicly available data\footnote{TokenTerminal provides a free premium account for three days upon registration.} and the steps to reproduce the results are well-described\footnote{The code is publicly available at \url{https://github.com/tengandreaxu/DeFi-value-investing}}. Thus, to the best of our knowledge, we are the first to provide a clear framework and steps to value \ac{DAO}\index{DAO} tokens and compare them with stocks of traditional publicly traded firms.
\end{enumerate}

This paper is organised as follows. In \autoref{sec:literature}, we provide a theoretical background. 
\autoref{sec:methodology} provides a description of our data set and technical details about the approaches used. 
\autoref{sec:results} shows our findings. 
Finally, \autoref{sec:conclusion} concludes.

\section{Literature}
\label{sec:literature}

To date, various valuation models have been developed from the literature \citep{damodaran2007valuation, damodaran2012investment}:

\begin{itemize}
    \item[Intrinsic] The theoretical value is a function of dividends, earnings, assets, liabilities, and so on. The theoretical value can be estimated through different approaches: \ac{DCF}, \ac{GGM} \citep{gordon1959dividends}, and the Modigliani \& Miller\index{Modigliani \& Miller} (MM) theorem \citep{modigliani1958cost}. 
    \item[Relative] This technique looks into {\em multiples} \citep{liu2002equity, schreiner2009equity} such as the Price to Sales (P/S) ratio.
\end{itemize}

We focus on \ac{DEFI}\index{DeFi} categories that have received major interest from the literature, that is, \ac{DEX}s\index{DEX} \citep{xu2021dexAmm}, \ac{PLF}s \citep{gudgeon2020DeFi}, and yield aggregators\index{aggregator} \citep{cousaert2022sok}. As \cite{damodaran2007valuation} reports, the majority of valuation approaches rely on \ac{DCF} and multiples. 

\section{Methodology}
\label{sec:methodology}

We apply both fundamental and comparable analyses to a set of \ac{DAO}\index{DAO} tokens in \ac{DEFI}\index{DeFi} and stocks in \ac{TradFi}\index{TradFi}. 
For \ac{DEFI}\index{DeFi}, we focus on \acp{DEX}\index{DEX}, \acp{PLF}, and yield aggregators\index{aggregator}, the three major \ac{DEFI}\index{DeFi} classes. The three analogous \ac{TradFi}\index{TradFi} sectors that we consider are exchanges\index{exchange}, banks, and asset managers, respectively. 
The set of token assets and firms grouped by their respective industry/financial service is shown in \autoref{tab:assets}.

\begin{table*}
\captionsetup{justification=centering}
\centering
\caption{\ac{DEFI} \ac{DAO} protocols and publicly traded companies universe examined.}
\scriptsize
\begin{tabular}{@{}rrrrcrrrcrrr@{}}\toprule
$\it{\ac{DEFI}}$ & \textbf{\acp{DEX}} && \textbf{\acp{PLF}} && \textbf{Yield aggregators} \\
\hphantom{$c$} & \acf{UNI}  && \acf{AAVE}  && \acf{YFI} \\
\hphantom{$c$} & \acf{CRV} && \acf{COMP} && \acf{IDLE} \\
\midrule
$\it{\ac{TradFi}}$ & \textbf{Exchanges} && \textbf{Banks} && \textbf{Asset managers} \\
\hphantom{$c$} & \acf{CBOE} && \acf{C} && \acf{BRK.B} \\
\hphantom{$c$} & \acf{NDAQ} && \acf{BAC} && \acf{MS} \\
\hphantom{$c$} & \acf{ICE} && \acf{WFC} &&  \acf{BLK} \\
\bottomrule
\end{tabular}
\label{tab:assets}
\end{table*}

\subsection{Data}
\label{sec:data}
We use Token Terminal\footnote{https://docs.tokenterminal.com/} for aggregated and processed financial metrics of tokens. In particular, we collect tokens' off-chain and on-chain \textit{daily} measures, for example, circulating market cap, price, \ac{TVL}, and protocol revenues. Protocol revenues, in this case, is defined as the amount of revenue that is distributed to token holders \citep{token_terminal_metrics}. We compare tokens' fundamentals to publicly traded companies balance sheets in the same industry.
For companies' quarterly fundamental information, we use common financial data sources such as \ac{CRSP} and Compustat, well-known databases in empirical asset pricing works, such as \cite{kelly2013market}, \cite{welch2008comprehensive} to name a few. Similar to tokens' metrics, we are interested in companies' market cap, share price, total revenue, net assets, and pre-tax income \citep{piq_wrds}. Note that for firm's and bank's total revenue, we use the Quarterly Total Revenue \citep{revtq_wrds} and Quarterly Total Current Operating Revenue \citep{tcoq_wrds} respectively. Finally, we obtain the firm's quarter net asset value by subtracting from the firm's quarter total asset value \citep{atq_wrds} the firm's quarter total debt \citep{dlcq_wrds}.

\subsection{Fundamental analysis}
\label{sec:fundamental_valuation}
We apply the \ac{DCF} approach for the fundamental analysis. \autoref{eq:dcf} shows how \ac{DCF} values firms according to their future cash flows discounted by a discount factor:

\begin{equation}
    \label{eq:dcf}
    \it{NPV} = \sum_{t=0}^n \frac{CASH_t}{(1+r)^t} + \frac{Terminal Value}{(1+r)^{t+1}},
\end{equation}
where the discount factor $r$ is usually identified as the \ac{WACC}. For tokens, $CASH$ is equivalent to the yearly protocol revenue, while for publicly traded firms, it corresponds to the pre-tax income. In this work, we will estimate a fair revenue growth of 5\% for the next five years. The \ac{WACC} is defined as:

\begin{equation}
\label{eq:wacc}
    \it{WACC} = \left(\frac{E}{E+D} \times R_e\right)  + \left(\frac{D}{E+D} \times R_d \times (1 - Tax Rate)\right),
\end{equation}
where $E$ is total equities, $D$ liabilities, $R_e$ the cost of capital, and $R_d$ the cost of debt. 

\begin{table*}
\centering
\label{table:wacc}
\caption{Variables used to compute firms' \ac{DCF}. We set $\it{MarketReturn}$ to the average yearly market return: 10\%. We use Damodaran's industry $\beta$ and $\it{TaxRate}$ \citep{damodaran_beta, damodaran_tax_rate}. The cost of debt $R_d$ was retrieved from the firms' 10-K document. The cost of capital $R_e$ is the product of $\beta$ and $\it{MarketReturn}$. Finally, we compute ${\it{WACC}}$ with \autoref{eq:wacc}.}
\scriptsize
\begin{tabular}{@{}rrrrrrcrrrrrr@{}}
\toprule
\phantom{abc} & \multicolumn{1}{c}{$\beta$} &
\phantom{abc} & \multicolumn{1}{c}{$Market Return$} &
\phantom{abc} & \multicolumn{1}{c}{$R_d$} &
\phantom{abc} & \multicolumn{1}{c}{$R_e$} &
\phantom{abc} & \multicolumn{1}{c}{$Tax Rate$} &
\phantom{abc} & \multicolumn{1}{c}{$WACC$} \\ \midrule

\textbf{BAC} & 1.12 && \textit{10\%} && \textit{2.85\%} && \textit{11.20\%} && \textit{14.69\%} && \textit{10.59\%} \\
\textbf{BLK} & 1.05 && \textit{10\%} && \textit{2.10\%} && \textit{10.50\%} && \textit{13.37\%} && \textit{10.46\%} \\
\textbf{BRK.B} & 1.05 && \textit{10\%} && \textit{3.30\%} && \textit{10.50\%} && \textit{13.37\%} && \textit{10.47\%} \\
\textbf{C} & 1.12 && \textit{10\%} && \textit{2.88\%} && \textit{11.20\%} && \textit{14.69\%} && \textit{10.08\%} \\
\textbf{CBOE} & 1.05 && \textit{10\%} && \textit{2.63\%} && \textit{10.50\%} && \textit{13.37\%} && \textit{10.47\%} \\
\textbf{ICE} & 1.12 && \textit{10\%} && \textit{3.00\%} && \textit{10.50\%} && \textit{13.37\%} && \textit{10.42\%} \\
\textbf{MS} & 1.12 && \textit{10\%} && \textit{2.90\%} && \textit{10.50\%} && \textit{13.37\%} && \textit{9.20\%} \\
\textbf{NDAQ} & 1.12 && \textit{10\%} && \textit{3.75\%} && \textit{10.50\%} && \textit{13.37\%} && \textit{10.28\%} \\
\textbf{WFC} & 1.05 && \textit{10\%} && \textit{2.37\%} && \textit{11.20\%} && \textit{14.69\%} && \textit{10.97\%} \\

\bottomrule
\end{tabular}
\end{table*}

For publicly traded firms, we estimate the cost of capital by multiplying the industry average $\beta$---retrieved from \cite{damodaran_beta}---with the historical market returns set to 10\%. The historical market return is coherent with the last 20 years’ S\&P 500 average annual return (9.87\%) \citep{damodaran_market_return}. The cost of debt---the interest rate that companies have to pay on their debt---is retrieved directly from the firm's 10-K filing from the \ac{SEC} Edgar database\footnote{https://www.sec.gov/edgar/searchedgar/companysearch.html}. 
Specifically, we use the companies' debt interest rate for senior structured notes stated in the \enquote{Financial Statements and Supplementary Data} section. Finally, the average corporate tax rate by industry is used \citep{damodaran_tax_rate}. 
By plugging these numbers into \autoref{eq:wacc}, we obtain the firm's \ac{WACC}. 
By contrary, we cannot use the same analysis to estimate \ac{WACC} for \ac{DEFI}\index{DeFi} tokens, as they achieve greater economy of scale and do not face corporate taxes. 
However, \ac{DEFI}\index{DeFi} tokens face higher risk, therefore, we discount their value by a higher factor---a value typically used in the literature: 25\% \citep{messari_discount}.
Next, $\it{TerminalValue}$ is simply defined as:

\begin{equation}
    \it{TerminalValue} = \sum_{i=1}^{\infty} \frac{\it{CASH}(1+ g)^n}{(1+r)^n} = \frac{\it{CASH}(1+ g)}{(r-g)},
\end{equation}
where $g$ is the perpetual growth rate, that is, the constant rate that a company is expected to grow at indefinitely. 
Commonly, $g$ should be in line with the nominal \ac{GDP} growth rate \citep{steiger2010validity}. 
We thus take the average yearly \ac{GDP} from 1990 to 2021 to proxy the perpetual growth rate: 2.39\% \citep{statista_gdp}. 
To conclude our assumptions, we estimate workforce expenses of 20\% and 30\% of the total income for \ac{DEFI}\index{DeFi} \ac{DAO}\index{DAO} tokens and publicly traded firms respectively. 
In \autoref{sec:results}, we show our estimations of growth for the next five years together with historical balance sheets. 
In these historical balance sheets, we report also the \ac{CQGR}, defined as:
\begin{equation}
    \label{eq:cqgr}
    \it{CQGR} = \left(\frac{V_{end}}{V_{start}}\right)^{\frac{1}{q}} - 1,
\end{equation}
where $q$ denotes the number of quarters between $start$ and ${end}$.
As the data for \ac{DEFI}\index{DeFi} is limited due to its short history, the analysis starts from the last quarter of 2020 for both \ac{DEFI}\index{DeFi} and \ac{TradFi}\index{TradFi}.

\subsection{Comparable Analysis}

We use multiples to compare valuation between \ac{DEFI}\index{DeFi} and \ac{TradFi}\index{TradFi}. 
Specifically, the \ac{DEFI}\index{DeFi} protocol/company's value is proxied by the token/stock' market cap. 
A firm's value is usually a function of the capacity of generating cash flow and risk, as reported in \cite{damodaran2007valuation}. Therefore, we identify the following:

\begin{itemize}
    \item[for \ac{DEFI}\index{DeFi}]
    \begin{itemize}
        \item \textit{revenue protocol}, as the protocol's total revenue. Since, TokenTerminal provides \textit{daily} observations, we simply sum up the daily revenue values within a quarter.
        \item \textit{treasury}, as the net asset owned by the underlying protocol, which in turn is collectively owned by protocol token holders.
    \end{itemize}
    \item[for \ac{TradFi}\index{TradFi}]
    \begin{itemize}
        \item \textit{revtq}, as \cite{revtq_wrds} reports, this value is the firm's quarterly total revenue.
        \item \textit{atq} and \textit{dlcq}, respectively, the firm's quarterly total assets and total liabilities \citep{atq_wrds, dlcq_wrds}. Therefore, the net assets owned by the firm is the difference between the two numbers.
    \end{itemize}
\end{itemize}

Hence, we end up with two multiples:

\begin{enumerate}
    \item The {\it Market Cap / Revenue} multiple: this ratio shows us how the underlying value changes with respect to its own income.
    \item The {\it Market Cap / Net Assets} multiple: similarly, this ratio shows us the change of value with respect to the net assets owned by the underlying firm/protocol.
\end{enumerate}

We show and discuss the results in the \autoref{sec:results}.

\section{Results}
\label{sec:results}

\subsection{Fundamental Analysis: Historical Balance Sheets}

In this section, we show the historical earnings for both \ac{DAO}\index{DAO} tokens and publicly traded firms. Due to \ac{DAO}\index{DAO} tokens' fairly recent phenomena, the historical data tables shown in \autoref{table:crypto_historical_data} and \autoref{table:firms_historical_data} start from Q4 2020. The tables show the underlying earnings from Q4 2020 to Q2 2022. In the last column, we compute and show the \ac{CQGR} defined in \autoref{eq:cqgr}.
The earnings, shown in USD millions, were steadily increasing for most \ac{DEFI}\index{DeFi} tokens until 2021's last quarter. However, with the recent bearish period within the crypto ecosystem, the slope has turned negative. Fears for inflation within risk markets, regulatory uncertainty, numerous attacks and hacks caused increasing diffidence and uncertainty in cryptoassets (see \cite{shanaev2020taming} and \cite{hu2020fluctuations}). This has caused investors to re-evaluate expectations and reprice the risk of \ac{DAO}\index{DAO} tokens. We can see that moving from 2022's first quarter to the second quarter, the growth is negative for all crypto-assets. By contrary, publicly traded companies have experienced mixed results. 
In fact, even though the COVID-19\index{COVID-19} crisis has increased volatility in the stock market, the 2022 decline seen for \ac{DEFI}\index{DeFi} tokens is less steep for publicly traded firms.
Unfortunately, the \ac{CQGR} numbers do not tell us much as we are considering different periods for \ac{DEFI}\index{DeFi} and \ac{TradFi}\index{TradFi} assets.
\begin{table*}\centering

\caption{\acp{DAO}' historical balance sheets.}
\label{table:crypto_historical_data}
\ra{1.3}
\scalebox{0.6}{
\begin{tabular}{@{}rrrrrrrrrrrrr@{}}\toprule
& \multicolumn{1}{c}{$2020$} &  
\phantom{abc} & \multicolumn{4}{c}{$2021$} &
\phantom{abc} & \multicolumn{2}{c}{$2022$} &
\phantom{abc} & \multicolumn{1}{c}{CQGR} \\
\cmidrule{2-2} \cmidrule{4-7}  \cmidrule{9-10}  \cmidrule{12-12}
& $Q4$ && $Q1$ & $Q2$ & $Q3$ & $Q4$ && $Q1$ & $Q2$ && \% \\ \midrule

\textbf{Uniswap}\\
Earnings (\$M) & $NA$ && 25.7 & 46.9 & 30.8 & 46.8 && 31.1 & 23.2 && \textit{-0.02\%}\\
\textit{\% growth} & \textit{$NA$} && \textit{$NA$} & \textit{82.33\%} & \textit{-34.33\%} & \textit{52.13\%} && \textit{-33.5\%} & \textit{-25.48\%} &&   \\

\textbf{Curve}\\
Earnings (\$M) & $NA$ && $NA$ & $NA$ & 6.34 & 19.8 && 18.5 & 12.2 && \textit{0.24\%} \\
\textit{\% growth} & \textit{$NA$} && \textit{$NA$} & \textit{$NA$} & \textit{$NA$} & \textit{213.04\%} && \textit{-6.78\%} & \textit{-34.21\%} &&   \\

\textbf{Compound}\\
Earnings (\$M) & $NA$ && 10.6 & 10.9 & 9.0 & 10.0 && 4.8 & 1.8 && \textit{-0.30\%} \\
\textit{\% growth} & 	\textit{$NA$} && \textit{$NA$} & \textit{2.50\%} & \textit{-16.91\%} & \textit{11.14\%} && \textit{-51.89\%} & \textit{-62.69\%} &&  \\

\textbf{AAVE}\\
Earnings (\$M) & $NA$ && 1.2 & 5.5 & 9.3 & 12.7 && 6.5 & 6.0 && \textit{0.37\%} \\
\textit{\% growth} & 	\textit{$NA$} && \textit{$NA$} & \textit{355.22\%} & \textit{68.66\%} & \textit{35.94\%} && \textit{-48.84\%} & \textit{-8.08\%} &&  \\

\textbf{Yearn Finance}\\
Earnings (\$M) & $NA$ && $NA$ & 22.3 & 23.1 & 26.2 && 14.0 & 6.9 && \textit{-0.25\%} \\
\textit{\% growth} & \textit{$NA$} && \textit{$NA$} & \textit{$NA$} & \textit{3.66\%} & \textit{12.98\%} && \textit{-46.55\%} & \textit{-50.29\%} &&   \\

\textbf{Idle Finance}\\
Earnings (\$M) & $NA$ && $NA$ & 0.3 & 0.2 & 0.2 && 0.1 & 0.05 && \textit{-0.37\%}\\
\textit{\% growth} & \textit{$NA$} && \textit{$NA$} & \textit{$NA$} & \textit{-36.96\%} & \textit{-24.12\%} && \textit{-59.15\%} & \textit{-20.77\%} &&   \\

\bottomrule
\end{tabular}
}
\end{table*}
\begin{table*}\centering

\caption{Firms' historical earnings.}
\label{table:firms_historical_data}
\ra{1.3}
\scalebox{0.6}{
\begin{tabular}{@{}rrrrrrrrrrrrr@{}}\toprule
& \multicolumn{1}{c}{$2020$} &  
\phantom{abc} & \multicolumn{4}{c}{$2021$} &
\phantom{abc} & \multicolumn{2}{c}{$2022$} &
\phantom{abc} & \multicolumn{1}{c}{CQGR} \\
\cmidrule{2-2} \cmidrule{4-7}  \cmidrule{9-10}  \cmidrule{12-12}
& $Q4$ && $Q1$ & $Q2$ & $Q3$ & $Q4$ && $Q1$ & $Q2$ && \% \\ \midrule

\textbf{ICE}\\
Earnings (\$M) & 583.0 && 674.0 & 833.0 & 1932.0 & 824.0 && 2109.0 & 832.0 && \textit{6.11\%} \\
\textit{\% growth} & \textit{$NA$} && \textit{15.61\%} & \textit{23.59\%} & \textit{131.93\%} & \textit{-57.35\%} && \textit{155.95\%} & \textit{-60.55\%} && \\

\textbf{NDAQ}\\
Earnings (\$M) & 356.0 && 267.0 & 395.0 & 454.0 & 371.0 && 314.0 & 374.0 && \textit{0.83\%} \\
\textit{\% growth} & \textit{$NA$} && \textit{-25.0\%} & \textit{47.94\%} & \textit{14.94\%} & \textit{-18.28\%} && \textit{-15.36\%} & \textit{19.11\%} &&  \\

\textbf{CBOE}\\
Earnings (\$M) & 163.4 && 122.9 & 192.9 & 179.2 & 173.8 && 210.2 & 224.9 && \textit{5.47\%} \\
\textit{\% growth} & \textit{$NA$} && \textit{-24.79\%} & \textit{56.96\%} & \textit{-7.1\%} & \textit{-3.01\%} && \textit{20.94\%} & \textit{6.99\%} &&   \\

\textbf{C} \\
Earnings (\$M) & 5,441.0 && 10,309.0 & 7,348.0 & 5,862.0 & 3,950.0 && 5,266.0 & 5,971.0 && \textit{6.07\%} \\
\textit{\% growth} & \textit{37.61\%} && \textit{89.47\%} & \textit{-28.72\%} & \textit{-20.22\%} & \textit{-32.62\%} && \textit{33.32\%} & \textit{13.39\%} &&      \\

\textbf{BAC}\\
Earnings (\$M) & 4,546.0 && 6,119.0 & 9,166.0 & 8,042.0 & 8,950.0 && 7,818.0 & 7,879.0 && \textit{9.6\%} \\
\textit{\% growth} & \textit{$NA$} && \textit{34.6\%} & \textit{49.8\%} & \textit{-12.26\%} & \textit{11.29\%} && \textit{-12.65\%} & \textit{0.78\%} &&    \\

\textbf{WFC}\\
Earnings (\$M) & 3,866.0 && 5,591.0 & 8,189.0 & 6,926.0 & 8,110.0 && 4,509.0 & 3,565.0 && \textit{1.03\%} \\
\textit{\% growth} & \textit{16.52\%} && \textit{44.62\%} & \textit{46.47\%} & \textit{-15.42\%} & \textit{17.1\%} && \textit{-44.4\%} & \textit{-20.94\%} &&     \\

\textbf{BRK.B} \\
Earnings (\$M) & 37,930.0 && 45,403.0 & 14,528.0 & 35,721.0 & 12,484.0 && 48,953.0 & 6,812.0 && \textit{-24.89\%} \\
\textit{\% growth} & \textit{$NA$} && \textit{19.7\%} & \textit{-68.0\%} & \textit{145.88\%} & \textit{-65.05\%} && \textit{292.13\%} & \textit{-86.08\%} &&       \\

\textbf{MS} \\
Earnings (\$M) & 4,430.0 && 5,344.0 & 4,566.0 & 4,874.0 & 4,884.0 && 4,588.0 & 3,319.0 && \textit{-0.7\%} \\
\textit{\% growth} & \textit{27.04\%} && \textit{20.63\%} & \textit{-14.56\%} & \textit{6.75\%} & \textit{0.21\%} && \textit{-6.06\%} & \textit{-27.66\%} &&       \\

\textbf{BLK} \\
Earnings (\$M) & 2,167.0 && 1,591.0 & 2,201.0 & 2,271.0 & 2,110.0 && 1,626.0 & 1,321.0 && \textit{-5.62\%} \\
\textit{\% growth} & \textit{9.39\%} && \textit{-26.58\%} & \textit{38.34\%} & \textit{3.18\%} & \textit{-7.09\%} && \textit{-22.94\%} & \textit{-18.76\%} &&       \\

\bottomrule
\end{tabular}
}
\end{table*}

\subsection{Fundamental Analysis: \ac{DCF} Estimation}
\begin{table*}\centering

\caption{\acp{DEX}' \ac{NPV} estimation using \ac{DCF}.}
\label{tab:dcf_dex}
\ra{1.3}
\scalebox{0.6}{
\begin{tabular}{@{}rrrrrrcrrrrrr@{}}\toprule
&  \multicolumn{1}{c}{$2022$} &
\phantom{abc} & \multicolumn{1}{c}{$2023$} &
\phantom{abc} & \multicolumn{1}{c}{$2024$} &
\phantom{abc} & \multicolumn{1}{c}{$2025$} &
\phantom{abc} & \multicolumn{1}{c}{$2026$} &
\phantom{abc} & \multicolumn{1}{c}{$2027$}\\ \midrule
\textbf{Uniswap}\\
Revenue (\$M)  &  108.68 && 114.12 && 119.82 && 125.81 && 132.10 && 138.71\\
Workforce expenses (\$M) & 21.73 && 22.82 && 23.96 && 25.16 && 26.42 && 27.74\\
Net income (\$M) & 86.95 && 91.3 && 95.86 && 100.65 && 105.69 && 110.97\\
PV cashflows (\$M) & 86.95 && 73.04 && 61.35 && 51.54 && 43.29 && 36.36\\
PV terminal value (\$M) &  && && && && && 131.74\\
\textbf{Total PV (\$M)} & 484.26\\
\textbf{Total PV / UNI supply (\$)} & 1.06\\
\textbf{UNI market price (\$)} & 5.00 \\
\midrule
\textbf{Curve}\\
Revenue (\$M)  &  61.31 && 64.37 && 67.59 && 70.97 && 74.52 && 78.25 \\
Workforce expenses (\$M) & 	12.26 && 12.87 && 13.51 && 14.19 && 14.90 && 15.65\\
Net income (\$M) & 49.05 && 51.5 && 54.08 && 56.78 && 59.62 && 62.6\\
PV cashflows (\$M) & 49.05 && 41.2 && 34.61 && 29.07 && 24.42 && 20.51\\
PV terminal value (\$M) &  && && && && && 74.31\\
\textbf{Total PV (\$M)} & 273.18\\
\textbf{Total PV / CRV supply (\$)} & 0.70\\
\textbf{CRV market price (\$)} & 0.69\\
\bottomrule
\end{tabular}
}
\newline\scriptsize
Note: market prices are retrieved on 30 June 2022. 
The annual growth rate for revenue and workforce expenses equals 5\% and 20\% respectively. 
\end{table*}
\begin{table*}\centering

\caption{\acp{PLF}' \ac{NPV} estimation using \ac{DCF}.}
\label{table:dcf_plf}
\ra{1.3}
\scalebox{0.6}{
\begin{tabular}{@{}rrrrrrcrrrrrr@{}}\toprule
&  \multicolumn{1}{c}{$2022$} &
\phantom{abc} & \multicolumn{1}{c}{$2023$} &
\phantom{abc} & \multicolumn{1}{c}{$2024$} &
\phantom{abc} & \multicolumn{1}{c}{$2025$} &
\phantom{abc} & \multicolumn{1}{c}{$2026$} &
\phantom{abc} & \multicolumn{1}{c}{$2027$}\\ \midrule
\textbf{Compound}\\
Revenue (\$M)  &  13.25 && 13.91 && 14.6 && 15.33 && 16.1 && 16.91 \\
Workforce expenses (\$M) & 	2.65 && 2.78 && 2.92 && 3.07 && 3.22 && 3.38\\
Net income (\$M) & 10.6 && 11.13 && 11.68 && 12.27 && 12.88 && 13.52\\
PV cashflows (\$M) & 10.6 && 8.9 && 7.48 && 6.28 && 5.28 && 4.43\\
PV terminal value (\$M) &  && && && && && 16.06\\
\textbf{Total PV (\$M)} & 59.02\\
\textbf{Total PV / COMP supply (\$)} & 8.59\\
\textbf{COMP market price (\$)} & 47.48 \\
\midrule
\textbf{AAVE}\\
Revenue (\$M)  &  24.93 && 26.18 && 27.48 && 28.86 && 30.3 && 31.82 \\
Workforce expenses (\$M) & 	4.99 && 5.24 && 5.5 && 5.77 && 6.06 && 6.36\\
Net income (\$M) & 19.94 && 20.94 && 21.99 && 23.09 && 24.24 && 25.45\\
PV cashflows (\$M) & 19.94 && 16.75 && 14.07 && 11.82 && 9.93 && 8.34\\
PV terminal value (\$M) &  && && && && && 30.22\\
\textbf{Total PV (\$M)} & 111.07\\
\textbf{Total PV / AAVE supply (\$)} & 8.01\\
\textbf{AAVE market price (\$)} & 57.00 \\
\bottomrule
\end{tabular}
}
\newline\scriptsize
Note: market prices are retrieved on 30 June 2022. 
The annual growth rate for revenue and workforce expenses equals 5\% and 20\% respectively. 
\end{table*}
\begin{table*}\centering

\caption{Yield aggregators' \ac{NPV} evaluation using \ac{DCF}.}
\label{tab:dcf_yearn}
\ra{1.3}
\scalebox{0.6}{
\begin{tabular}{@{}rrrrrrcrrrrrr@{}}\toprule
&  \multicolumn{1}{c}{$2022$} &
\phantom{abc} & \multicolumn{1}{c}{$2023$} &
\phantom{abc} & \multicolumn{1}{c}{$2024$} &
\phantom{abc} & \multicolumn{1}{c}{$2025$} &
\phantom{abc} & \multicolumn{1}{c}{$2026$} &
\phantom{abc} & \multicolumn{1}{c}{$2027$}\\ \midrule

\textbf{Yearn Finance}\\
Revenue (\$M)  &  41.87 && 43.96 && 46.16 && 48.47 && 50.9 && 53.44 \\
Workforce expenses (\$M) & 	8.37 && 8.79 && 9.23 && 9.69 && 10.18 && 10.69\\
Net income (\$M) & 33.5 && 35.17 && 36.93 && 38.78 && 40.72 && 42.75\\
PV cashflows (\$M) & 33.5 && 28.14 && 23.64 && 19.85 && 16.68 && 14.01\\
PV terminal value (\$M) &  && && && && && 50.75\\
\textbf{Total PV (\$M)} & 	186.56\\
\textbf{Total PV / YFI supply (\$)} & 5,902.42\\
\textbf{YFI market price (\$)} & 5,419.10 \\
\midrule
\textbf{Idle Finance}\\
Revenue (\$M)  &  0.25 && 0.26 && 0.27 && 0.28 && 0.3 && 0.31 \\
Workforce expenses (\$M) & 	0.05 && 0.05 && 0.05 && 0.06 && 0.06 && 0.06\\
Net income (\$M) & 0.2 && 0.21 && 0.22 && 0.23 && 0.24 && 0.25\\
PV cashflows (\$M) & 0.2 && 0.16 && 0.14 && 0.12 && 0.1 && 0.08\\
PV terminal value (\$M) &  && && && && && 0.30\\
\textbf{Total PV (\$M)} & 1.10\\
\textbf{Total PV / IDLE supply (\$)} & 0.41\\
\textbf{IDLE market price (\$)} & 0.23 \\
\bottomrule
\end{tabular}
}
\newline\scriptsize
Note: market prices are retrieved on 30 June 2022. 
The annual growth rate for revenue and workforce expenses equals 5\% and 20\% respectively. 
\end{table*}
Following the historical balance sheets, we show the results of the \ac{DCF} estimation with assumptions that are described in \autoref{table:crypto_historical_data}. Surprisingly, the projections show that a number of \ac{DEFI}\index{DeFi} tokens are relatively \textit{hyped}. As shown in \autoref{tab:dcf_dex} and \autoref{table:dcf_plf}, \ac{UNI}, \ac{AAVE}, and \ac{COMP} exhibit overvaluation with prices at least five-fold of their theoretical level. 
By contrast, \autoref{tab:dcf_yearn} shows that yield aggregators\index{aggregator} are undervalued relative to their fundamentals. 
Yield aggregators\index{aggregator} use smart contracts that implement algorithmic investment strategies with their earnings mostly coming from performance fees of their vaults \citep{Xu2023AProtocols}. 
In this context, a vault is another name for a smart contract which allocates locked assets and algorithmically takes portfolio\index{portfolio} management decisions. It is interesting to see how both \ac{DEFI}\index{DeFi} and \ac{TradFi}\index{TradFi} asset management protocols and firms are undervalued compared to their theoretical valuations.
For valuation of public firms shown in \autoref{table:cex_valuation}, \autoref{table:cbanks_valuation}, \autoref{table:c_yield_valuation}, apparently overpriced relative to their fundamentals, NDAQ\index{NDAQ} and BLK\index{BLK} have experienced a recent downwards correction from the market \citep{bloomberg_blk_loss, ft_ndaq_loss}. 
It's interesting to point out that albeit tokens' overvaluation, hedge funds, such as BlackRock or Grayscale, and banks, JP Morgan or Citigroup for example, are investing in crypto-assets. 



\begin{table*}\centering

\caption{\acp{CEX}' \ac{NPV} evaluation using \ac{DCF}.}
\label{table:cex_valuation}
\ra{1.3}
\scalebox{0.6}{
\begin{tabular}{@{}rrrrrrcrrrrrr@{}}\toprule
&  \multicolumn{1}{c}{$2022$} &
\phantom{abc} & \multicolumn{1}{c}{$2023$} &
\phantom{abc} & \multicolumn{1}{c}{$2024$} &
\phantom{abc} & \multicolumn{1}{c}{$2025$} &
\phantom{abc} & \multicolumn{1}{c}{$2026$} &
\phantom{abc} & \multicolumn{1}{c}{$2027$}\\ \midrule

\textbf{Intercontinental Exchange}\\
Revenue (\$M)  &  5,882.0 && 6,176.1 && 6,484.9 && 6,809.15 && 7,149.61 && 7,507.09 \\
Workforce expenses (\$M) & 	1,764.6 && 1,852.83 && 1,945.47 && 2,042.74 && 2,144.88 && 2,252.13\\
Net income (\$M) & 4,117.4 && 4,323.27 && 4,539.43 && 4,766.4 && 5,004.73 && 5,254.96\\
PV cashflows (\$M) & 4,117.4 && 3,915.12 && 3,722.77 && 3,539.88 && 3,365.97 && 3,200.61\\
PV terminal value (\$M) &  && && && && && 40,667.48\\
\textbf{Total PV (\$M)} & 62,529.23\\
\textbf{Total PV / ICE supply (\$)} & 111.86\\
\textbf{ICE market price (\$)} & 94.04 \\
\midrule
\textbf{Nasdaq}\\
Revenue (\$M)  &  1,376.0 && 1,444.8 && 1,517.04 && 1,592.89 && 1,672.54 && 1,756.16 \\
Workforce expenses (\$M) & 	412.8 && 433.44 && 455.11 && 477.87 && 501.76 && 526.85\\
Net income (\$M) & 963.2 && 1,011.36 && 1,061.93 && 1,115.02 && 1,170.78 && 1,229.31\\
PV cashflows (\$M) & 963.2 && 917.06 && 873.14 && 831.32 && 791.5 && 753.59\\
PV terminal value (\$M) &  && && && && && 9,720.70\\
\textbf{Total PV (\$M)} & 14,850.51\\
\textbf{Total PV / NDAQ supply (\$)} & 90.28\\
\textbf{NDAQ market price (\$)} & 152.54 \\
\midrule

\textbf{Chicago Board of Trade}\\
Revenue (\$M)  &  870.2 && 913.71 && 959.4 && 1,007.36 && 1,057.73 && 1,110.62 \\
Workforce expenses (\$M) & 	261.06 && 274.11 && 287.82 && 302.21 && 317.32 && 333.19\\
Net income (\$M) & 609.14 && 639.6 && 671.58 && 705.16 && 740.41 && 777.43 \\
PV cashflows (\$M) & 609.14 && 578.98 && 550.32 && 523.07 && 497.17 && 472.56\\
PV terminal value (\$M) &  && && && && && 5,976.65\\
\textbf{Total PV (\$M)} & 9,207.89\\
\textbf{Total PV / CBOE supply (\$)} & 86.71\\
\textbf{CBOE market price (\$)} & 113.19 \\
\bottomrule
\end{tabular}
}
\newline\scriptsize
Note: market prices are retrieved on 30 June 2022. 
The annual growth rate for revenue and workforce expenses equals 5\% and 30\% respectively. 
\end{table*}
\begin{table*}\centering

\caption{Banks' \ac{NPV} evaluation using \ac{DCF}.}
\label{table:cbanks_valuation}
\ra{1.3}
\scalebox{0.6}{
\begin{tabular}{@{}rrrrrrcrrrrrr@{}}\toprule
&  \multicolumn{1}{c}{$2022$} &
\phantom{abc} & \multicolumn{1}{c}{$2023$} &
\phantom{abc} & \multicolumn{1}{c}{$2024$} &
\phantom{abc} & \multicolumn{1}{c}{$2025$} &
\phantom{abc} & \multicolumn{1}{c}{$2026$} &
\phantom{abc} & \multicolumn{1}{c}{$2027$}\\ \midrule

\textbf{Citigroup}\\
Revenue (\$M)  &  522,474.0 && 23,597.7 && 24,777.58 && 26,016.46 && 27,317.29 && 28,683.15 \\
Workforce expenses (\$M) & 	16,742.2 && 7,079.31 && 7,433.28 && 7,804.94 && 8,195.19 && 8,604.95\\
Net income (\$M) & 15,731.8 && 16,518.39 && 17,344.31 && 18,211.53 && 19,122.1 && 20,078.21\\
PV cashflows (\$M) & 15,731.8 && 15,006.1 && 14,313.88 && 13,653.59 && 13,023.76 && 12,422.99\\
PV terminal value (\$M) &  && && && && && 163,802.15\\
\textbf{Total PV (\$M)} & 247,954.28\\
\textbf{Total PV / C supply (\$)} & 128.02\\
\textbf{C market price (\$)} & 46.02 \\
\midrule

\textbf{Bank of America}\\
Revenue (\$M)  &  31,394.0 && 32,963.7 && 34,611.88 && 36,342.48 && 38,159.6 && 40,067.58 \\
Workforce expenses (\$M) & 	9,418.2 && 9,889.11 && 10,383.56 && 10,902.74 && 11,447.88 && 12,020.28\\
Net income (\$M) & 21,975.8 && 23,074.59 && 24,228.32 && 25,439.74 && 26,711.72 && 28,047.31\\
PV cashflows (\$M) & 21,975.8 && 20,864.52 && 19,809.44 && 18,807.71 && 17,856.64 && 16,953.66\\
PV terminal value (\$M) &  && && && && && 211,690.06\\
\textbf{Total PV (\$M)} & 327,957.84\\
\textbf{Total PV / BAC supply (\$)} & 40.68\\
\textbf{BAC market price (\$)} & 31.15 \\
\midrule

\textbf{Well Fargo \& Co.}\\
Revenue (\$M)  &  16,148.0 && 16,955.4 && 17,803.17 && 18,693.33 && 19,628.0 && 20,609.4 \\
Workforce expenses (\$M) & 	4,844.4 && 5,086.62 && 5,340.95 && 5,608.0 && 5,888.4 && 6,182.82\\
Net income (\$M) & 11,303.6 && 11,868.78 && 12,462.22 && 13,085.33 && 13,739.6 && 14,426.58 \\
PV cashflows (\$M) & 11,303.6 && 10,695.06 && 10,119.27 && 9,574.49 && 9,059.03 && 8,571.33\\
PV terminal value (\$M) &  && && && && && 102,944.03\\
\textbf{Total PV (\$M)} & 162,266.81\\
\textbf{Total PV / WFC supply (\$)} & 42.78\\
\textbf{WFC market price (\$)} & 39.17 \\
\bottomrule
\end{tabular}
}
\newline\scriptsize
Note: market prices are retrieved on 30 June 2022. 
The annual growth rate for revenue and workforce expenses equals 5\% and 30\% respectively. 
\end{table*}
\begin{table*}
\centering
\caption{Asset management firms' \ac{NPV} evaluation using \ac{DCF}.}
\label{table:c_yield_valuation}
\ra{1.3}
\scalebox{0.60}{
\begin{tabular}{@{}rrrrrrcrrrrrr@{}}\toprule
&  \multicolumn{1}{c}{$2022$} &
\phantom{abc} & \multicolumn{1}{c}{$2023$} &
\phantom{abc} & \multicolumn{1}{c}{$2024$} &
\phantom{abc} & \multicolumn{1}{c}{$2025$} &
\phantom{abc} & \multicolumn{1}{c}{$2026$} &
\phantom{abc} & \multicolumn{1}{c}{$2027$}\\ \midrule

\textbf{Berkshire Hathaway}\\
Revenue (\$M)  &  111,530.0 && 117,106.5 && 122,961.82 && 129,109.92 && 135,565.41 && 142,343.68 \\
Workforce expenses (\$M) & 	33,459.0 && 35,131.95 && 36,888.55 && 38,732.98 && 40,669.62 && 42,703.1 \\
Net income (\$M) & 78,071.0 && 81,974.55 && 86,073.28 && 90,376.94 && 94,895.79 && 99,640.58\\
PV cashflows (\$M) & 78,071.0 && 74,205.71 && 70,531.78 && 67,039.76 && 63,720.62 && 60,565.82\\
PV terminal value (\$M) &  && && && && && 765,995.62\\
\textbf{Total PV (\$M)} & 1,180,130.30\\
\textbf{Total PV / BRK.B supply (\$)} & 535.06\\
\textbf{BRK.B market price (\$)} & 273.02 \\
\midrule

\textbf{Morgan Stanley}\\
Revenue (\$M)  &  15,814.0 && 16,604.7 && 17,434.94 && 18,306.68 && 19,222.02 && 20,183.12 \\
Workforce expenses (\$M) & 4,744.2 && 4,981.41 && 5,230.48 && 5,492.0 && 5,766.6 && 6,054.94\\
Net income (\$M) & 11,069.8 && 11,623.29 && 12,204.45 && 12,814.68 && 13,455.41 && 14,128.18\\
PV cashflows (\$M) & 11,069.8 && 10,644.02 && 10,234.61 && 9,840.96 && 9,462.44 && 9,098.48 \\
PV terminal value (\$M) &  && && && && && 132,466.85\\
\textbf{Total PV (\$M)} & 192,817.15\\
\textbf{Total PV / MS supply (\$)} & 111.91\\
\textbf{MS market price (\$)} & 76.10 \\
\midrule

\textbf{BlackRock}\\
Revenue (\$M)  &  5,894.0 && 6,188.7 && 6,498.14 && 6,823.04 && 7,164.19 && 7,522.4  \\
Workforce expenses (\$M) & 	41,768.2 && 1,856.61 && 1,949.44 && 2,046.91 && 2,149.26 && 2,256.72\\
Net income (\$M) & 4,125.8 && 4,332.09 && 4,548.7 && 4,776.13 && 5,014.94 && 5,265.68 \\
PV cashflows (\$M) & 4,125.8 && 3,921.81 && 3,727.91 && 3,543.59 && 3,368.39 && 3,201.84\\
PV terminal value (\$M) &  && && && && && 40,528.04\\
\textbf{Total PV (\$M)} & 62,417.382\\
\textbf{Total PV / BLK supply (\$)} & 413.45\\
\textbf{BLK market price (\$)} & 609.04 \\
\bottomrule
\end{tabular}
}
\newline\scriptsize
Note: market prices are retrieved on 30 June 2022. 
The annual growth rate for revenue and workforce expenses equals 5\% and 30\% respectively. 
\end{table*}

\subsection{Comparable Analysis}

In this section, we discuss the results of our comparable analysis. In the comparable analysis, we define multiples and compare different assets of the same kind: exchanges\index{exchange}, banking, and asset managers. 
Due to their nascency, \ac{DEFI}\index{DeFi} protocols have fewer data points than their \ac{TradFi}\index{TradFi} counterparties. 
We show that the multiples change over time starting from Q1 2021 to Q2 2022. 

\begin{figure}[t]
    \centering
    \subfloat[\acp{DEX} and exchanges \label{fig:dex_revenue}]{
        \includegraphics[width=0.3\linewidth]{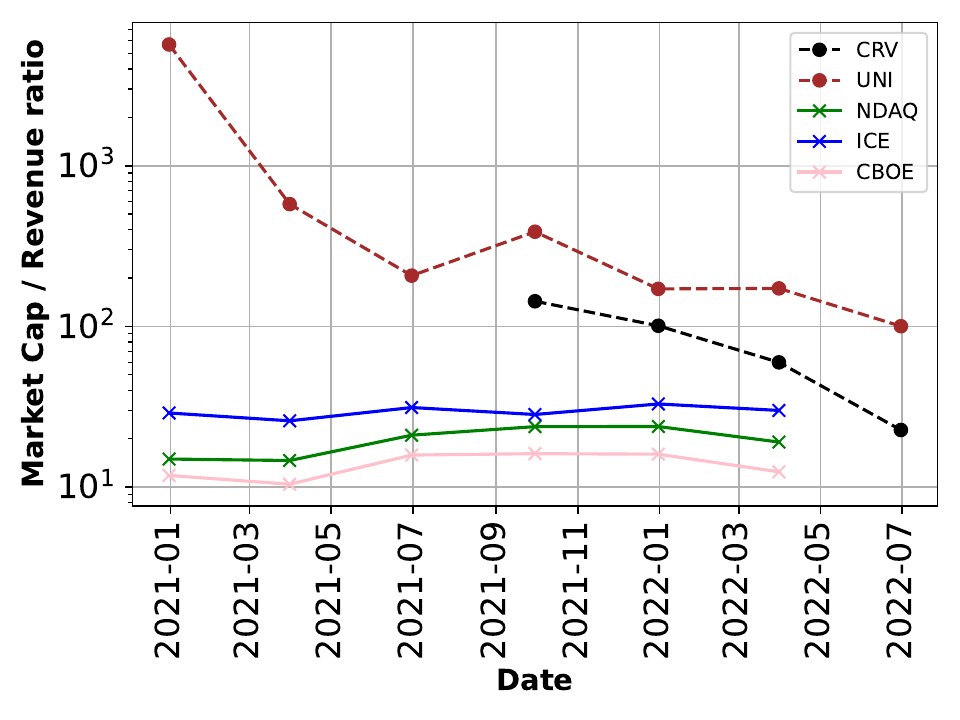}
    }
    \subfloat[\acp{PLF} and Banks \label{fig:plf_revenue}]{
        \includegraphics[width=0.3\linewidth]{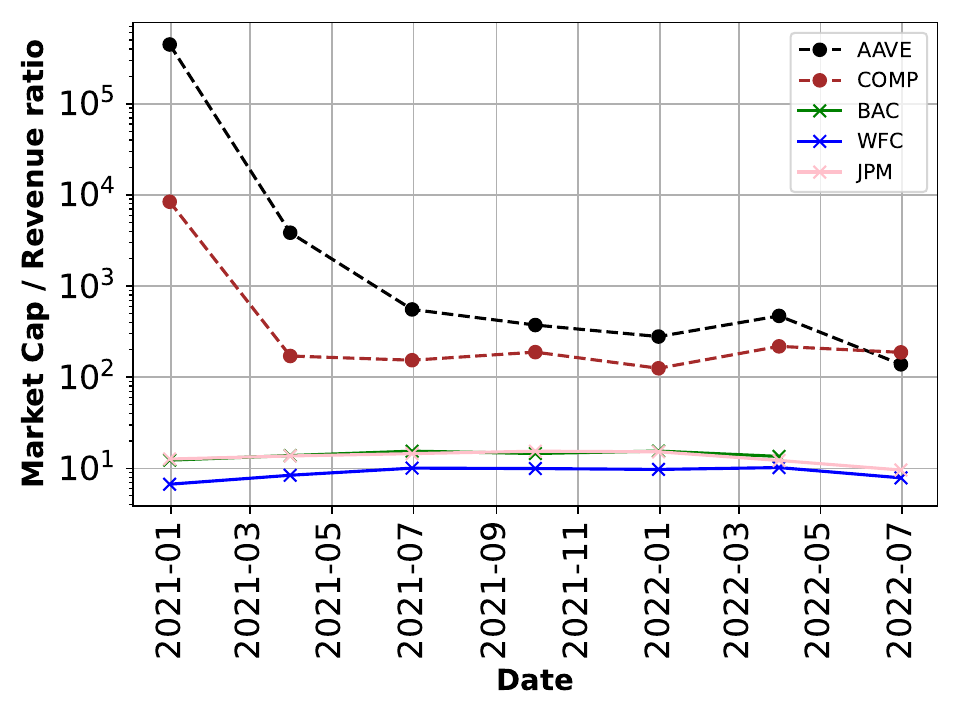}
    }
    \subfloat[Yield aggregators and asset managers \label{fig:yield_agg_revenue}]{
        \includegraphics[width=0.3\linewidth]{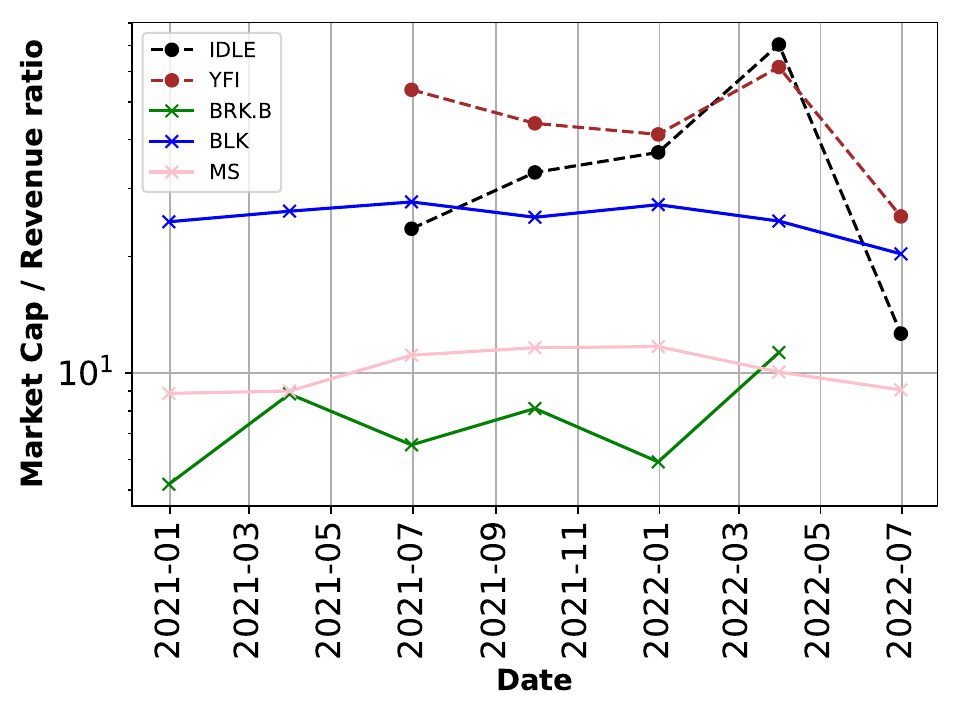}
    }
    \caption{The historical (log) {\it Market Cap /  Revenue} ratio.} 
    \label{fig:revenue}
\end{figure}

\autoref{fig:revenue} show the {\it Market Cap / Revenue} ratio for (a) exchanges\index{exchange}, (b) banks, and (c) asset managers. We merge both \ac{DEFI}\index{DeFi} and \ac{TradFi}\index{TradFi} assets. These figures show that \ac{DEFI}\index{DeFi} assets were relatively overvalued with respect to their \ac{TradFi}\index{TradFi} counterparties, and have been only recently converging to a more fair price. 
In fact, in Q2 2022, we can see \ac{DEFI}\index{DeFi} converging to \ac{TradFi}\index{TradFi} multiples values. This could be explained by the slump in the crypto market where high growth expectations have been adjusted towards more realistic levels. 
Due to the large price increases in crypto tokens and high growth in \ac{DEFI}\index{DeFi} protocol usage, aggressive (growth) expectations were priced in. 
The crypto market slump and sluggish recovery caused a reassessment of the aggressive growth expectations and the subsequent revaluation of \ac{DEFI}\index{DeFi} tokens. As previously discussed, comparable analysis assumes that market prices reflect investors expectations on average. 
Therefore, by comparing similar assets' multiples, the empirical observations can reveal undervalued and/or overvalued assets. 
For example, by looking at \autoref{fig:dex_revenue}, Uniswap\index{Uniswap} during 2021 was overvalued compared to its \ac{TradFi}\index{TradFi} counterparties and only recently its price has been converging to a \enquote{fair} price.

\begin{figure}[t]
    \centering
    \subfloat[\acp{DEX} and exchanges]{
        \includegraphics[width=0.3\linewidth]{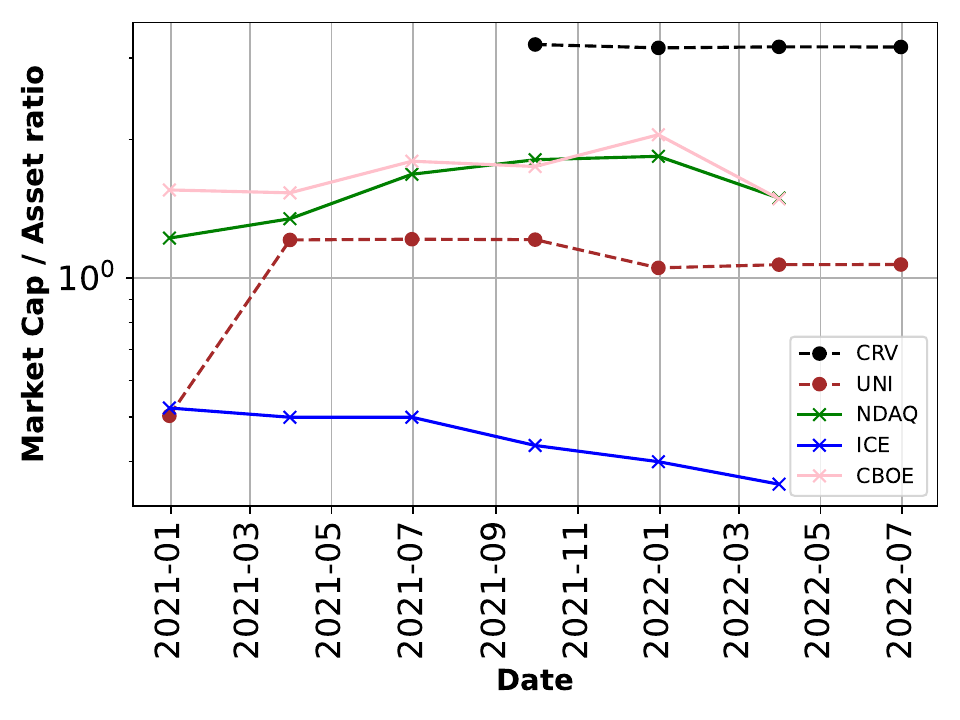}
         \label{fig:dex_net_assets}
    }
    \subfloat[\acp{PLF} and Banks \label{fig:plf_net_assets}]{
        \includegraphics[width=0.3\linewidth]{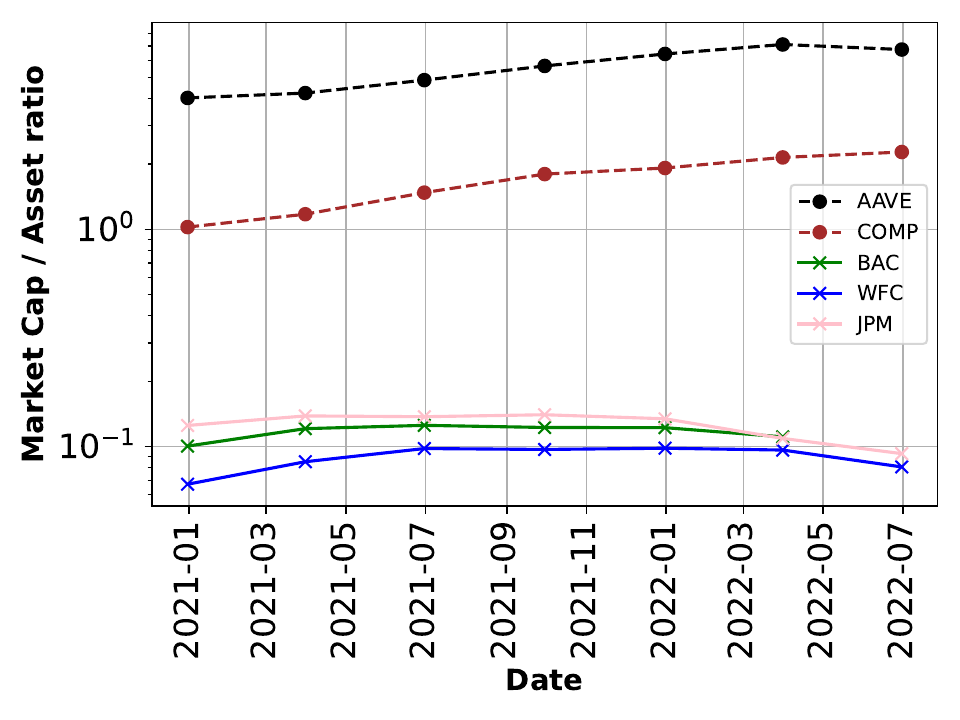}
    }
    \subfloat[Yield aggregators and asset managers \label{fig:yield_agg_net_assets}]{
        \includegraphics[width=0.3\linewidth]{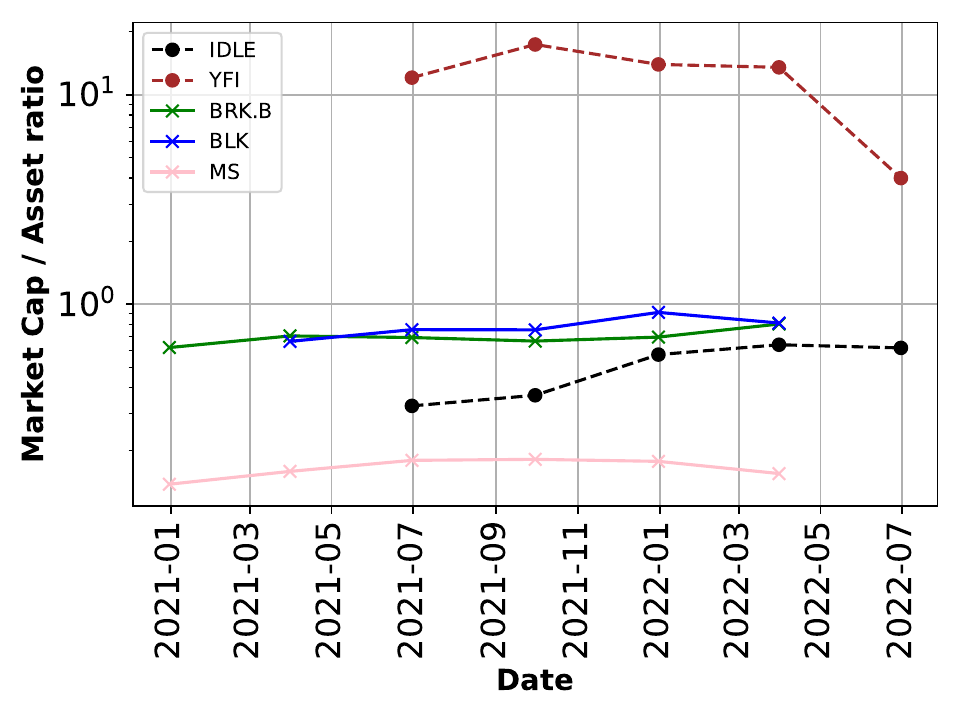}
    }
    \caption{Historical (log) {\it Market Cap / Net Asset} ratio.} 
    
    \label{fig:net_asset}
\end{figure}

Similarly, \autoref{fig:net_asset} shows the {\it Market Cap / Net Assets} ratio for the same set of financial services and assets from \autoref{fig:revenue}. Net assets are represented by the protocol treasury value for \ac{DEFI}\index{DeFi}, and equal the difference between total assets and total liabilities for \ac{TradFi}\index{TradFi}. 
We generally observe overvaluation for \ac{DEFI}\index{DeFi} tokens which is, differently from the latest multiples, not disappearing with the recent market slump. 
It should be noted that this behaviour is mixed between different \ac{DEFI}\index{DeFi} tokens. 
First, in \autoref{fig:dex_net_assets}, we can see that Uniswap\index{Uniswap} is \enquote{fairly} valued with respect to its \ac{TradFi}\index{TradFi} equivalents. 
This could be explained by Uniswap\index{Uniswap} retaining 430M UNI tokens in its own treasury that can be distributed to the community in the future \citep{uni_treasury}. 
To date, these tokens have a value of ~\$3B, and Uniswap\index{Uniswap}'s market cap is around ~\$5B. 
This relatively large treasury size makes Uniswap\index{Uniswap} the \enquote{richest} \ac{DEFI}\index{DeFi} protocol with the highest treasury value \citep{uni_treasury_cryptofees}. Compared to Uniswap\index{Uniswap}, other protocols' treasury values have been driven by crypto-assets that users pay as fees for using the \ac{DEFI}\index{DeFi} service \citep{Xu2023AProtocols}. 
Therefore, moving to \autoref{fig:plf_net_assets}, we can observe that---since the slopes are positive---the recent market cap crash was lower than their treasury loss. 
Finally, \autoref{fig:yield_agg_net_assets} shows an overpriced \ac{YFI} consistent with \autoref{fig:yield_agg_revenue}. Interestingly, Idle Finance's {\it Market Cap / Net Assets} ratio is comparable to \ac{TradFi}\index{TradFi} Hedge Funds\index{fund}.

\section{Conclusion}
\label{sec:conclusion}

In this study,
we apply conventional valuation analysis on \ac{DEFI}\index{DeFi} tokens and provide a comparison with the valuation on stocks of publicly listed firms within similar categories. 
Specifically, we analyse \acp{DEX}\index{DEX}, \acp{PLF}, and yield aggregators\index{aggregator}, which are also compared with exchanges\index{exchange}, banks and asset managers, respectively. 

The results show that \ac{DEFI}\index{DeFi} tokens have been rather overpriced relative to the equity of financial services firms. More specifically, \ac{DEFI}\index{DeFi} assets were overvalued with respect to their \ac{TradFi}\index{TradFi} counterparties, and have been only recently converging to a more fair price. We believe high growth expectations in \ac{DEFI}\index{DeFi} have been adjusted towards more realistic levels in line with \ac{TradFi}\index{TradFi}. 

Despite the large growth in \ac{DEFI}\index{DeFi} tokens and cryptocurrencies\index{cryptocurrency} in general, little research has been conducted on the valuation of these. Research on digital assets and \ac{DEFI}\index{DeFi} tokens valuation is still in an early stage, and we contribute by providing a framework to think about fundamental valuation in \ac{DEFI}\index{DeFi}. 
Future research could build upon our approach by further developing fundamental measures and frameworks speciﬁc to \ac{DEFI}\index{DeFi}. 

\chapterbib

\end{document}